# Optimal Denial-of-Service Attacks Against Status Updating

Saad Kriouile, Mohamad Assaad
Laboratory of Signals and Systems (L2S)
CentraleSupelec, University of Paris-Saclay
Emails: {saad.kriouile, mohamad.assaad}@centralesupelec.fr

Deniz Gündüz, Touraj Soleymani
Department of Electrical and Electronic Engineering
Imperial College London
Emails: {d.gunduz, touraj}@imperial.ac.uk

*Abstract*—In this paper, we investigate denial-of-service attacks against status updating. The target system is modeled by a Markov chain and an unreliable wireless channel, and the performance of status updating in the target system is measured based on two metrics: age of information and age of incorrect information. Our objective is to devise optimal attack policies that strike a balance between the deterioration of the system's performance and the adversary's energy. We model the optimal problem as a Markov decision process and prove rigorously that the optimal jamming policy is a threshold-based policy under both metrics. In addition, we provide a low-complexity algorithm to obtain the optimal threshold value of the jamming policy. Our numerical results show that the networked system with the age-of-incorrect-information metric is less sensitive to jamming attacks than with the age-of-information metric.

## I. INTRODUCTION

Cyber-physical systems are complex systems that tightly integrate computational algorithms with dynamical processes, providing a seamless interaction between the digital and physical worlds [2]. These systems leverage computation, communication, and control to enhance efficiency, adaptability, and automation in various modern domains in our societies such as smart cities, smart factories, smart healthcare, and smart transportation. However, it is essential to acknowledge the dynamic nature of cyber-physical systems, which necessitates persistent monitoring and status updating with the purpose of capturing latest information about the physical environment. This influx of real-time data empowers cyber-physical systems to respond swiftly to changing conditions, ensuring that decisions are grounded in the most current and relevant information [1], [3], [4], [11], [13].

Consider scenarios such as autonomous vehicles navigating through traffic or smart grids managing energy distribution. In these instances, real-time decision-making relies on up-to-the-moment status updates. For autonomous vehicles, the system must constantly analyze sensory measurements to make fast decisions about navigation, adapting to the road conditions and traffic patterns. Similarly, in smart grids, sensory measurements enable the system to balance energy supply and demand dynamically, optimizing resource allocation based on the current state of the grid. In essence, real-time decision-making in cyber-physical systems hinges on the currency and accuracy of status updates.

Nevertheless, cyber-physical systems are highly vulnerable to cyber attacks [5]. Three common types of cyber attacks on cyber-physical systems are [6]: deception attacks, eavesdropping attacks, and denial-of-service attacks. In broad terms, deception attacks compromise data integrity by manipulating data, eavesdropping attacks compromise data confidentiality by intercepting data, whereas denial-of-service attacks, which are the primary focus of our study, compromise data availability by obstructing its transmission in networks. In the present paper, we look at the problem of status updating of a simple cyber-physical system, with a sensor and a monitor that are connected through a lossy communication channel. This networked system operates in the presence of an adversary with the intent of executing denial-of-service attacks. We examine the problem from the perspective of the adversary, which is able to jam the channel at each time instant. Our objective is to devise optimal attack policies that strike a balance between the deterioration of the system's performance and the adversary's energy.

In this paper, the performance of status updating is captured based on two metrics: Age of Information (AoI) and Age of Incorrect Information (AoII). AoI is a metric introduced for the first time in [7] to capture the freshness of information in real-time monitoring systems. Since its introduction, this metric has attracted the attention of researchers [1], [8]–[11]. However, AoI is independent of the contents of messages and it increases even if the status of the source has not changed. To overcome this is issue, AoII was proposed in [12], [13], which takes into account both the change of the source status and the freshness of information. The readers are referred to [14]–[17], and the references therein, for more details on these metrics and their use for real-time monitoring and cyber-physical systems.

### A. Literature Survey

There exists a body of research on the design of denial-of-service attack policies in the context of cyber-physical systems [18]–[24]. Notably, in this class, Zhang *et al.* [18] obtained the optimal signal-independent jamming blockage policy[1] that maximizes a regulation loss function subject to a blockage frequency constraint in a networked control system, and extended the results to a multiple subsystem case where the adversary should select one channel at each time. Zhang *et al.* [19] also derived the optimal signal-independent jamming blockage policy that maximizes a distortion loss function subject to a blockage frequency constraint in a networked estimation system, and extended the results when there exists a packet-drop-ratio stealthiness constraint. Qin *et al.* [20] found the optimal signal-independent jamming blockage policy that maximizes a distortion loss function subject to a blockage frequency constraint in a networked estimation system when the channel is subject to packet loss even in the absence of attacks, and extended the results to a multiple subsystem case where the adversary should select only one channel at each time. Zhang *et al.* [21] obtained the optimal signal-independent jamming power policy[2] that maximizes a distortion loss function subject to a jamming power constraint in a networked estimation system when the channel is ideal in the absence of attacks, and presented their findings for both static and dynamic attacks. Qin *et al.* [22] obtained the optimal signal-independent jamming power policy that maximizes a distortion loss function subject to a jamming power constraint in a networked estimation system when the channel is subject to packet loss even in the absence of attacks, and provided their results for both static and dynamic attacks. Gan *et al.* [23] found the optimal signal-independent jamming power policy that maximizes a distortion loss function subject to a jamming power constraint in a networked estimation system when the sensor is connected to the monitor via a two-hop relay channel and the adversary should select one of the hops at each time. In addition, Zhang *et al.* [24] obtained the optimal signal-dependent jamming power policy that maximizes a regulation loss function subject to a soft jamming power constraint in a networked control system when the channel is fading, proposed a suboptimal attack policy, and extended the results to different packet detection schemes.

---
[1]In this case, the adversary's decision at each time is either to block the channel or not to block it.

To the best of our knowledge, only few works addressed the problem of denial-of-service attacks in the area of the AoI [25]–[27]. In [25], a continuous-time system model is considered with a reliable channel. The attacker aims to find a jamming time distribution that maximizes the AoI, and the interaction between the attacker and the source is formulated as a dynamic game. In [26], the interaction between the attacker and the source is rather considered at the physical layer and it is formulated as a static game while considering the channel between the source and destination as an M/G/1/1 queue. In [26], a continuous-time M/M/1 model between the source and the destination is considered, and the interaction between the attacker and the source is also formulated as a game problem. Our work differs from the aforementioned studies, since we consider both the AoI and AoII metrics, and adopt a discrete-time source model with an unreliable channel between the source and the monitor. We consider also that if the adversary decides to attack the channel, it succeeds with a given probability, which represents the fact that the channel of the jammer is not perfect and may suffer from deep fading. Finally, our work considers also that the adversary has limited energy.

---

[2]In this case, the adversary's decision at each time is to choose the amount of the jamming power.

### B. Overview and Organization

In this paper, we investigate denial-of-service attacks against status updating. The target system is modeled by a Markov chain and an unreliable wireless channel, and the performance of status updating in the target system is measured based on two metrics: AoI and AoII. Our objective is to devise optimal attack policies that strike a balance between the deterioration of the system's performance and the adversary's energy. We model the optimal problem as a Markov Decision Process (MDP) and prove rigorously that the optimal jamming policy is a threshold-based policy under both metrics. In addition, we provide a low-complexity algorithm to obtain the optimal threshold value of the jamming policy. Our numerical results show that the networked system with the AoII metric is less sensitive to jamming attacks than with the AoI metric.

The paper is organized as follows. We present the system model and state the problem of interest in Section II. We provide our main theoretical results on the design of optimal jamming policies in Section III. We provide our numerical results in Section IV. Finally, we conclude the paper and discuss potential future research directions in Section V.

## II. SYSTEM MODEL AND PROBLEM STATEMENT

In this section, we present the system model and state the problem of interest. We start first by describing the system model and defining the parameters of the system namely, the channel state, the successful attack's probabilities and the source's transition probability. Then, we present the metrics dynamics in our system model, specifically, we describe the evolution of the AoI and AoII metrics. Finally, we present an average reward maximization problem studied in this paper.

### A. Networked System Description

We consider in our paper a binary Markovian source that generates and sends status updates about the process of interest to a monitor over an unreliable independent and identically distributed (i.i.d.) channel. Time is discretized and normalized to the time slot duration. The attacker aims to prevent the source from delivering correctly the packet to the monitor by jamming the channel. We let $c(t)$ refer to the channel realization at time slot $t$: if $c(t) = 1$, then the transmitted packet is successfully decoded by the receiver side and $c(t) = 0$ otherwise. The probability that $c(t) = 1$ equals $p$. On the other hand, the attacker which plays the role of the agent in our system model has two possible actions at time $t$: $a(t) = 1$ if the agent attacks the channel, and $a(t) = 0$ otherwise. When the attacker decides to attack the channel, the probability that it succeeds in jamming it is $q$. Regarding the binary source's parameter, we consider that the probability to move to a different state is $r$. Similar to [15] , we assume in this paper that $r < 1/2$. We consider also that the transmission is instantaneous, i.e., the information will be transmitted and eventually received by the monitor within the same time slot.

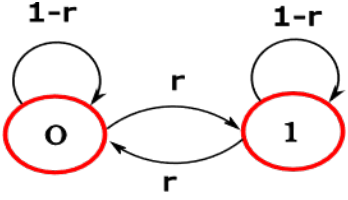

Fig. 1: A binary Markovian source.

### B. Metrics Evolution

*1) AoI Metric Evolution:* We describe how AoI evolves in our system settings. For that, we let $s(t)$ be the value of AoI at time $t$. The AoI is incremented by one if $c(t) = 0$ or if the agent performs a successful attack on the channel; and goes to 0 otherwise. Therefore, the transition probabilities are as follows [3]:

- $\Pr\big(s(t+1) = s(t)+1 | a(t+1) = 1\big) = q + (1-q)(1-p)$.
- $\Pr\big(s(t+1) = s(t)+1 | a(t+1) = 0\big) = 1-p$.

*2) AoII Metric Evolution:* We describe how AoII evolves in our system settings. We let $s(t)$ be the value of AoII at time $t$. We distinguish between two cases of $s(t)$ ($s(t) = 0$ or $s(t) > 0$). If $s(t) = 0$, this means that the estimated state at the side of the monitor equals to that of the source, then, if the agent decides to jam the channel at time slot $t+1$, AoII will be incremented by one in two cases: 1) if the source moves to the next state and $c(t) = 0$, 2) if the source moves to the next state and the $c(t) = 1$ but the agent succeeds to jam the channel; otherwise AoII goes to 0. However, if the agent decides to stay idle, AoII will be incremented by one if the source moves to the next state and $c(t) = 0$. On the other hand, when $s(t) > 0$, this means that the estimated state is different from that of the source, then, if the agent decides to attack the channel, AoII will be incremented by one in two cases: 1) if the source remains at the same state and $c(t) = 0$, 2) if the source remains at the same state and $c(t) = 1$ but the agent succeeds to jam it; otherwise it goes to 0. However, if the agent decides to stay idle, then AoII is incremented by one if the source remains at the same state and $c(t) = 0$. Therefore, the transition probabilities are as follows:

- $\Pr\big(s(t+1) = s(t)+1 | s(t) = 0, a(t+1) = 1\big) = pqr + (1-p)r$.
- $\Pr\big(s(t+1) = s(t)+1 | s(t) = 0, a(t+1) = 0\big) = (1-p)r$.
- $\Pr\big(s(t+1) = s(t)+1 | s(t) > 0, a(t+1) = 1\big) = pq(1-r) + (1-p)(1-r)$.
- $\Pr\big(s(t+1) = s(t)+1 | s(t) > 0, a(t+1) = 0\big) = (1-p)(1-r)$.

---

[3] We omit the transition probabilities to the state 0 since $s(t)$ can move either to $s(t)+1$ or 0, then the sum of the two transition probabilities equals 1

### C. Problem Statement

The decision maker in our problem is the attacker. We suppose that if the attacker decides to jam the channel, an additional positive cost, i.e., negative reward, is incurred due to the energy consumed during the attacking process. Let $R(t) = s(t) - \lambda a(t)$ be the reward function of the agent at time slot $t$, where $\lambda > 0$. A jamming policy $\phi$ is defined as a sequence of actions $\phi = (a^\phi(0), a^\phi(1), \ldots)$, where $a^\phi(t) = 1$ if the agent decides to jam the channel at time $t$, and $a^\phi(t) = 0$ otherwise. Our aim is to find an optimal jamming policies that maximizes the expected average reward. This can be formulated as follows:

$$\underset{\phi \in \Phi}{\text{maximize}} \quad \lim_{T \to +\infty} \sup \frac{1}{T} \mathbb{E}^{\phi \in \Phi}\Big(\sum_{t=0}^{T-1} R^\phi(t) | s(0)\Big). \quad (1)$$

where $\Phi$ denotes the set of all causal jamming policies. We characterize the optimal solutions to this problem in the next section.

## III. OPTIMAL JAMMING POLICIES

In this section, we provide our main theoretical results on the design of optimal jamming policies. We start first by showing that the optimal solution is threshold-based policy. Then, we provide a closed-form expression of the problem of interest. Exploiting that, we derive the optimal threshold value and we present a low-complexity algorithm that finds the optimal threshold as a function of $\lambda$.

### A. Structural Results

The optimization problem in (1) can be viewed as an infinite-horizon average-reward MDP problem with the following characteristics:

- *State*: The state of the MDP at time $t$ is the AoI function or AoII function denoted by $s(t)$.
- *Action*: The action at time $t$, denoted by $a(t)$, specify if the channel is attacked (value 1) or not (value 0).
- *Transition probabilities*: The transition probabilities specify the probabilities associated with state changes given actions.
- *Reward*: The instantaneous reward of the MDP, $R(s(t), a(t))$, which is equal to $s(t) - \lambda a(t)$.

The optimal policy $\phi^*$ of Problem (1) can be obtained by solving the following Bellman equation for each state $s$:

$$\theta + V(s) = \max_{a \in \{0,1\}} \Big\{ s - \lambda a + \sum_{s' \in \mathbb{N}} \Pr(s \to s'|a) V(s') \Big\} \quad (2)$$

where $\Pr(s \to s'|a)$ is the transition probability from state $s$ to $s'$ under action $a$, $\theta$ is the optimal value of the problem, $V(s)$ is the differential reward-to-go function, and $\mathbb{N}$ is the set of positive integers.

**Theorem 1.** *The optimal solution of the AoI-based and AoII-based maximization problems in (2) is an increasing threshold-based policy. Explicitly, there exists $n \in \mathbb{N}$ such that when the current state $s < n$, the prescribed action is a passive action, and when $s \geq n$, the prescribed action is an active action.*

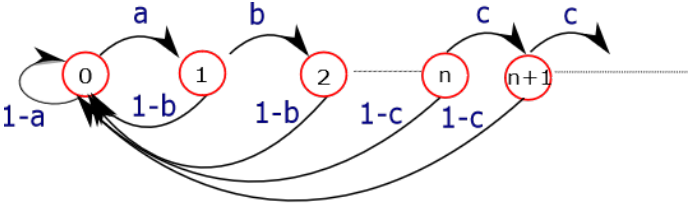

Fig. 2: The state transition under a threshold policy with parameter $n$ for a generic system model

*Proof:* The proof can be found in Appendix A. ∎

According to Figure 2, and bearing in mind the system parameters described in Section II-A, for the AoI-based model, we have $a = b = 1-p$ and $c = q+(1-p)(1-q)$, and for the AoII-based model, we have $a = (1-p)r$, $b = (1-p)(1-r)$ and $c = pq(1-r)+(1-p)(1-r)$.

The results of Theorem 1 can be interpreted as follows. When the state is less than the optimal threshold, the impact of the attacker's action on the reward is small, whereas this action's impact becomes more important when the state is bigger than the optimal threshold value. Accordingly, the attacker saves energy until the state is large enough to reach the optimal threshold, which happens due to the unreliable channel, and then jams the channel and keeps doing this as far as the state is larger than the optimal threshold value.

### B. Closed-form Expression of the Optimal Solution

We have proved that the optimal solution of (1) is a threshold-based policy. Nevertheless, we still have to determine the optimal threshold value for any given $\lambda$. To that end, we will first derive a simple closed-form expression for the average reward. Note that we can derive the steady-state form of the problem in (1) under a given threshold policy with parameter $n$ as

$$\underset{n \in \mathbb{N}}{\text{maximize}} \quad \overline{s^n} - \lambda \overline{a^n} \quad (3)$$

where $\overline{s^n}$ is the average value of AoI or AoII, and $\overline{a^n}$ is the average active time under the threshold policy with parameter $n$. Specifically:

$$\overline{s^n} = \lim_{T \to +\infty} \sup \frac{1}{T} \mathbb{E}^n \Big( \sum_{t=0}^{T-1} s(t) | s(0), \phi^t(n) \Big) \quad (4)$$

$$\overline{a^n} = \lim_{T \to +\infty} \sup \frac{1}{T} \mathbb{E}^n \Big( \sum_{t=0}^{T-1} a(t) | s(0), \phi^t(n) \Big) \quad (5)$$

where $\phi^t(n)$ denotes the threshold policy with parameter $n$.

To compute $\overline{s^n}$ and $\overline{a^n}$, we will determine the stationary distribution of the Discrete Time Markov Chain (DTMC) that models the evolution of the AoI and the AoII under the threshold policy with parameter $n$. To have a comprehensive analysis, we consider a more general DTMC, as illustrated in Figure 2. As a result, the derived stationary distribution will be expressed in terms of the parameters $a$, $b$, and $c$ for enhanced generality. This broader scope allows us to obtain analytical expressions that are applicable across a range of scenarios, capturing the nuanced dynamics of the system.

**Proposition 1.** *For a given threshold with parameter $n$, the DTMC admits $u_n(i)$ as its stationary distribution:*

$$u_n(i) = \begin{cases} \frac{(1-b)(1-c)}{(1-b+a)(1-c)+(c-b)b^{n-1}} & \text{if } i = 0 \\ ab^{i-1} \frac{(1-b)(1-c)}{(1-b+a)(1-c)+(c-b)b^{n-1}} & \text{if } 1 \leq i \leq n \\ ab^{n-1} c^{i-n} \frac{(1-b)(1-c)}{(1-b+a)(1-c)+(c-b)b^{n-1}} & \text{if } i \geq n+1 \end{cases} \quad (6)$$

*Proof:* See Appendix B. ∎

By leveraging the above results, we can now proceed with finding a closed-form expression for the average reward under the threshold policy with parameter $n$.

**Proposition 2.** *Under a threshold policy $n$, the average reward denoted by $\overline{s^n}$ is equal to:*

$$\overline{s^n} = \frac{a(1-c)[1-(n+1)b^n+nb^{n+1}]}{(1-b)[(1-b+a)(1-c)+ab^{n-1}(c-b)]} + \frac{a(1-b)b^{n-1}c[n(1-c)+1]}{(1-c)[(1-b+a)(1-c)+ab^{n-1}(c-b)]} \quad (7)$$

*Proof:* By exploiting the results of Proposition 1 and by the definition of $\overline{s^n}$ given in (4), we get after algebraic manipulations the desired results. ∎

Now, we derive the average active time. For that, we exploit the results of Proposition 1 and use the definition of $\overline{a^n}$ given in (5).

**Proposition 3.** *The average active time denoted by $\overline{a^n}$ is equal to*

$$\overline{a^n} = \begin{cases} 1 & \text{if } n = 0 \\ \frac{a(1-b)b^{n-1}}{(1-b+a)(1-c)+a(c-b)b^{n-1}} & \text{if } n > 0 \end{cases} \quad (8)$$

*Proof:* By exploiting the results in Proposition 1 and the expression (5), we obtain the desired results. ∎

Now that we derived the steady-state form of the problem in (1), our goal is to find the threshold value $n$ that maximizes $s^n - \lambda a^n$. A brute-force scheme computes $s^n - \lambda a^n$ for different threshold values $n \in \mathbb{N}$. However, this process will be endless since as AoI and AoII evolve within an infinite state space. To overcome this issue, we develop a low-complexity algorithm that enables us to determine the optimal threshold value for any given $\lambda$.

The next definition defines the sequence $\lambda(n)$.

**Definition 1.** *$\lambda(n)$ is the intersection point between $\overline{s^n} - \lambda \overline{a^n}$ and $\overline{s^{n+1}} - \lambda \overline{a^{n+1}}$, i.e.,*

$$\lambda(n) = \frac{\overline{s^{n+1}} - \overline{s^n}}{\overline{a^{n+1}} - \overline{a^n}} \quad (9)$$

The next theorem provides the optimal threshold value as a function of $\lambda$.

**Theorem 2.** *The optimal threshold value policy of the problem in* (1) *satisfies:*
- *If $\lambda \leq \lambda(0)$, then the optimal threshold is $0$*
- *If $\lambda(n) < \lambda \leq \lambda(n+1)$, then the optimal threshold is $n+1$*

*Proof.* See Appendix C. □

According to the above theorem and the algorithmic developments in [28], we provide Algorithm 1 that enables us to find the optimal threshold value.

---
**Algorithm 1** Optimal Threshold Policy
---
1: Input $\lambda$ and $\lambda(\cdot)$.
2: Init. $t = 0$, $x_0 = 0$ and $k = 1$
3: **if** $\lambda \leq \lambda(0)$ **then**: $n^* = 0$
4: **else**
5:     **while** $k == 1$ **do** $x_{t+1} = x_t + \alpha(\lambda - \lambda(x_t))$
6:         **if** $\lambda(\lfloor x_{t+1} \rfloor) < \lambda \leq \lambda(\lfloor x_{t+1} \rfloor + 1)$ **then**
7:             $k = 0$ and $n^* = \lfloor x_{t+1} \rfloor + 1$
8:         **end if**
9:     **end while**
10: **end if**
11: **return** $n^*$

---

In Algorithm 1, $\lfloor x \rfloor$ represents the integer part of $x$, and use the continuous extension of the function $\lambda(.)$ in $\mathbb{R}^+$ based on the linear interpolation, i.e.,

$$\lambda(x) = \begin{cases} \lambda(i) & \text{if } x = i \in \mathbb{N} \\ \lambda(i+1)(x-i) & \\ \quad -\lambda(i)(x-i-1) & \text{if } x \in [i, i+1] \end{cases} \quad (10)$$

## IV. NUMERICAL RESULTS

In this section, we compare the performance of the optimal policies in the AoI and AoII models with the uniform random solution and Opposite-optimal policies. To do so, we let the AoI-optimal and AoII-optimal policies be the optimal threshold-based policies that maximize the average reward for the AoI-based and AoII-based models, respectively. Moreover, the uniform random policy is the policy that jams the channel with probability $1/2$ at every time slot. The Opposite-AoI-optimal and Opposite-AoII-optimal policies consist of selecting the opposite action of the AoI-optimal and AoII-optimal policies respectively. In more details, if the AoI (or AoII) is less than the optimal threshold value the attacker jams the channel, and it stays silent otherwise. These opposite policies are based on the intuition that the attacker may be attempted to attack the channel when the age metric is small in order to increase it. Recall that $\lambda$ represents the energy cost term, or equivalently the amount of the energy consumed by the attacker when it decides to jam the channel. In our simulations, we would like to illustrate the evolution of the the average reward with the AoI metric under the AoI-optimal and random policies as a function of $\lambda$, and also the evolution the average reward with the AoII metric under the AoII-optimal and random policies.

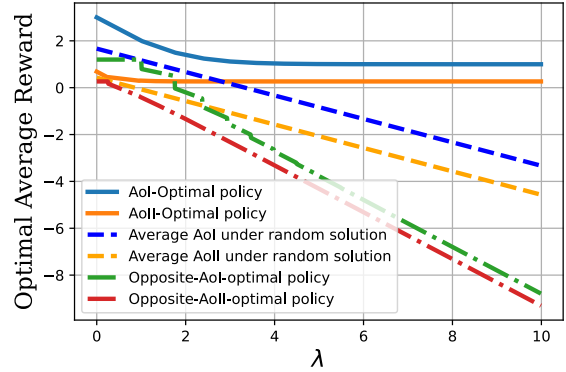

Fig. 3: Comparison between AoI-optimal and AoII-optimal policies in term of average reward under the first scenario

We suppose that $\lambda$ varies from 1 to 10 with step size 0.1, and we consider the following parameters $p = q = 1/2$, $r = 1/4$. In Figure 3, we observe that the average reward under the AoI-optimal solution decreases as $\lambda$ grows and converges to a fixed value. Indeed, when $\lambda$ increases, the cost of jamming the channel becomes more substantial in terms of energy which reduces significantly the performance of the optimal solution. Moreover, we observe that the attacker performs poorly in term of maximizing the average reward with the AoII metric even with small $\lambda$ compared to the AoI metric. The reason lies in the lack of control the attacker has over the source state. Specifically, if the estimated state at the monitoring side is accurate, jamming the channel becomes ineffective as AoII remains at the state 0. On the contrary, deciding to jam the channel increases the probability of AoI being incremented by one. We also observe that for both metrics, the uniform random solution is sub-optimal, which corroborate our theoretical results. Lastly the performance of the opposite-optimal policies highlight the fact that the intuitive solution that corresponds to attack the channel when the value of the AoI or AoII is small is sub-optimal. This shows that the optimal solution is not trivial and illustrates further the importance of our theoretical results.

## V. CONCLUSION

This paper considered the problem of denial-of-service attacks against status updating. The target system was modeled by a Markov chain and an unreliable wireless channel, and the performance of status updating in the target system was measured based on two metrics: AoI and AoII. We provided a rigorous analysis, and derived the structures of the optimal jamming policies that achieve a trade-off between the deterioration of the system's performance and the attacker's energy. More specifically, we proved that the optimal jamming policy is a threshold-based policy, and developed a low-complexity algorithm to find the optimal threshold value. The numerical results showed that jamming has a higher impact on the networked system with the AoII metric than with the AoI metric.

## VI. Supplementary numerical results

In this section, we compare between the performance of the optimal policies in the AoI and AoII models for different scenarios.

In our simulations, we would like to illustrate the effect of the channel parameter $p$ on the performance of AoI-optimal and AoII-optimal solutions. In addition, we would like to illustrate the optimal threshold values for the AoI-optimal and AoII-optimal polices as functions of $\lambda$. For that we consider these two scenarios: scenario 1) $p = q = 1/2$, $r = 1/4$, scenario 2) $p = 0.1$, $q = 1/2$ and $r = 1/2$.

According to Figure 4, the optimal threshold of both policies increases with respect to $\lambda$. This is due to the fact that as $\lambda$ grows, the attacker will consume more energy when it attacks the channel, then to compensate this increase of energy, it will reduces the average active time by increasing the threshold policy. We should emphasize that our numerical results in Figures 3 and 4 are compatible with our theoretical results. In particular, we observe that for large values of $\lambda$, the optimal threshold values are considerably large, which implies that the corresponding reward is converging according to Proposition 2.

Regarding the second scenario when $p$ is relatively small, the plots in Figures 5 and 6 compared to that of Figures 3 and 4 (first scenario), show us clearly that the performance of AoI-policy is remarkably improved while that of AoII-policy is slightly improved. This phenomenon is linked to the situation where $p$ is small. Even with only a few channel attacks, the average age tends to be high due to a significant probability of packet failure in reaching its destination. In contrast, AoII is less susceptible to these effects compared to AoI. This is because the evolution of AoII relies not only on the channel state but also on the state of the Markovian source, providing a higher degree of resilience in the face of adverse conditions.

## Appendix A
## Proof of theorem 1

We start first by providing the explicit expression of the Bellman equation for both AoI-based and AoII-based maxi-mization problem. The explicit expression of Bellman equation at state $s$ for AoI-based maximization problem:

$$\theta + V(s) = \max \{s + pV(0) + (1-p)V(s+1);$$
$$s - \lambda + (q + (1-p)(1-q))V(s+1) + (1-q)pV(0)\} \quad (11)$$

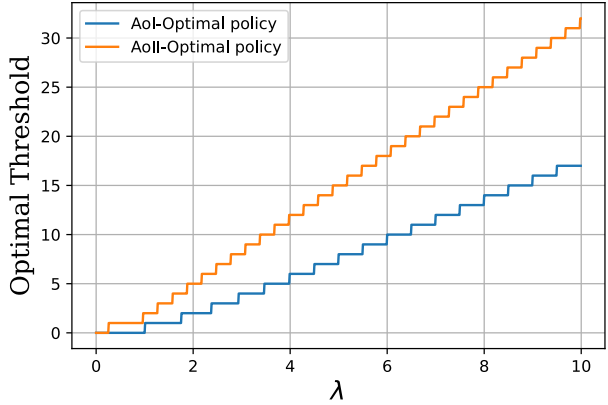

Fig. 4: Comparison between AoI-optimal and AoII-optimal policies in term of threshold value under the first scenario

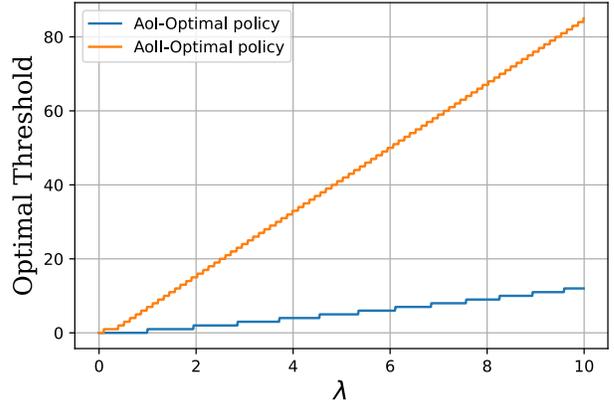

Fig. 6: Comparison between AoI-optimal and AoII-optimal policies in term of threshold value under the second scenario

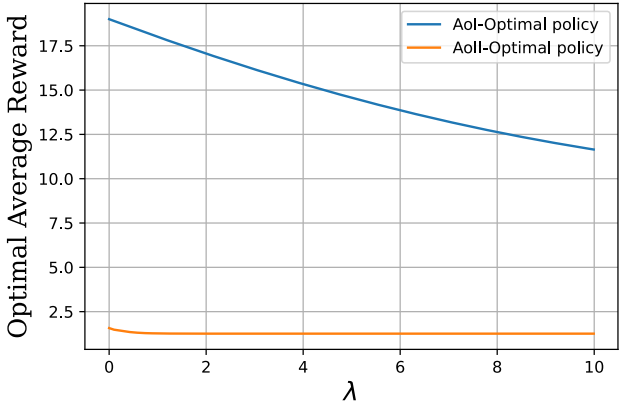

Fig. 5: Comparison between AoI-optimal and AoII-optimal policies in term of average reward under the second scenario

The explicit expression of Bellman equation at state $s$ for AoII-based maximization problem:

- $s = 0$:

$$\theta + V(s) = \max \{s + (1-p)rV(s+1)$$
$$+ (p + (1-p)(1-r))V(0);$$
$$s - \lambda + (pqr + (1-p)r)V(s+1)$$
$$+ (1 - pqr - (1-p)r)V(0)\} \quad (12)$$

- $s > 0$:

$$\theta + V(s) = \max \{s + (1-p)rV(s+1)$$
$$+ (p + (1-p)r)V(0);$$
$$s - \lambda + (pq(1-r) + (1-p)(1-r))V(s+1)$$
$$+ (1 - pq(1-r) - (1-p)(1-r))V(0)\} \quad (13)$$

For the sake of space, we prove the present theorem for one of the two metrics. We consider for that the AoII metric since the proof requires relatively more analysis as the expression of the Bellman equation at the state $s = 0$ is different than that at the state $s > 0$. For that, we adopt the same approach used in [12] that consists of proving that the Value function is increasing using the Relative Value Iteration equation, then deducing that the optimal solution is threshold increasing policing by establishing that $\Delta V(s) = V^1(s) - V^0(s)$ is increasing with the state $s$ where $V^1(s)$ and $V^0(s)$ are the Value function evaluated at state $s$ if the action is $1$ and $0$ respectively.

**Lemma 1.** $V(.)$ *is an increasing function with $s$.*

*Proof:* The Relative Value Iteration equation consists of updating the value function $V^t(.)$ as follows:

$$V_{t+1}(s) = \max \{V_t^0(s), V_t^1(s)\} \quad (14)$$

where:

- if $s = 0$:

$$V_t^0(s) = s + (1-p)rV_t(s+1)$$
$$+ (p + (1-p)(1-r))V_t(0)$$
$$V_t^1(s) = s - \lambda + (pqr + (1-p)r)V_t(s+1)$$
$$+ (1 - pqr - (1-p)r)V_t(0) \quad (15)$$

- if $s > 0$:

$$V_t^0(s) = s + (1-p)(1-r)V_t(s+1)$$
$$+ (p + (1-p)r)V_t(0)$$
$$V_t^1(s) = s - \lambda + (pq(1-r) + (1-p)(1-r))V_t(s+1)$$
$$+ (1 - pq(1-r) - (1-p)(1-r))V_t(0) \quad (16)$$

Exploiting this equation above, we prove the present lemma by induction. In fact, we show that $V_t(\cdot)$ is increasing for all

$t$ and we conclude for $V(\cdot)$.

As $V_0(.) = 0$, then the property holds for $t = 0$. If $V_t(.)$ is increasing and with $s$, we show that, $V_{t+1}^0(s) \leq V_{t+1}^0(s+1)$ and $V_{t+1}^1(s) \leq V_{t+1}^1(s+1)$. First, we establish that $V_{t+1}(.)$ is increasing when $s \geq 1$.

$$V_{t+1}^0(s+1) - V_{t+1}^0(s) = 1 + (1-p)(1-r)(V_t(s+1) - V_t(s)) \tag{17}$$

Since $V_t(.)$ is increasing with $s$, then: $V_{t+1}^0(s+1) - V_{t+1}^0(s) \geq 0$.

As consequence, $V_{t+1}^0(\cdot)$ is increasing with $s$.

In the same way, we have:
$$\begin{aligned}V_{t+1}^1(s+1) - V_{t+1}^1(s) \\ = 1 + \big(pq(1-r) \\ + (1-p)(1-r)\big)(V_t(s+1) - V_t(s))\end{aligned} \tag{18}$$

Hence:
$$V_{t+1}^1(s+1) - V_{t+1}^1(s) \geq 0 \tag{19}$$

As consequence, $V_{t+1}^1(\cdot)$ is increasing with $s$.
Since $V_{t+1}(.) = \max\{V_{t+1}^0(\cdot), V_{t+1}^1(\cdot)\}$, then $V_{t+1}(.)$ is increasing with $s$ when $s \geq 1$. To prove that $V_{t+1}(.)$ is increasing with $s$ in $\mathbb{N}$, we should show that $V_{t+1}(0) \leq V_{t+1}(1)$. We have:

$$\begin{aligned}V_{t+1}^0(1) - V_{t+1}^0(0) &= 1 + (1-p)(1-r)V_t(2) \\ &\quad - (1-p)rV_t(1) - (1-2r)(1-p)V_t(0) \\ &\stackrel{(a)}{\geq} 1 + (1-p)(1-2r)V_t(1) \\ &\quad - (1-p)(1-2r)V_t(0) \\ &\stackrel{(a)}{\geq} 1 + (1-p)(1-2r)(V_t(1) - V_t(0)) \\ &\stackrel{(b)}{\geq} 0\end{aligned} \tag{20}$$

where $(a)$ comes from the fact that $V_t(.)$ is increasing with $s$ and $(b)$ comes from the fact that $r \leq 1/2$. Following the same steps, we show that $V_{t+1}^1(1) - V_{t+1}^1(0) \geq 0$. Hence, $V_{t+1}(0) \leq V_{t+1}(1)$. As consequence, $V_{t+1}(\cdot)$ is increasing with $s$ in $\mathbb{N}$. Accordingly, we demonstrated by induction that $V_t(.)$ is increasing for all $t$. Knowing that $\lim_{t \to +\infty} V_t(s) = V(s)$, $V(.)$ is also increasing with $s$. ∎

We define:
$$\Delta V(s) = V^1(s) - V^0(s) \tag{21}$$

where $\lim_{t \to +\infty} V_t^0(s) = V^0(s)$ and $\lim_{t \to +\infty} V_t^1(s) = V^1(s)$.
Subsequently, $\Delta V(s)$ equals to:
- if $s > 0$
$$\Delta V(s) = -\lambda + (1-p)(1-r)(V(s+1) - V(0)) \tag{22}$$
- if $s = 0$
$$\Delta V(0) = -\lambda + (1-p)r(V(1) - V(0)) \tag{23}$$

According to Lemma 1, $V(.)$ is increasing with $s$. Therefore, $\Delta V(s)$ is increasing with $s$ for $s > 0$. Given that $r \leq 1/2$, then $\Delta V(0) \leq -\lambda + (1-p)(1-r)(V(1) - V(0)) \leq \Delta V(s)$ for all $s > 0$. Thus, $\Delta V(s)$ is increasing with $s$ in $\mathbb{N}$. Hence, there exists $n$ such that for all $s < n$, $\Delta V(s) \leq 0$, and for all $s \geq n$, $\Delta V(s) \geq 0$. Given that the optimal action for state $s$ is the one that maximizes $\{V^0(\cdot), V^1(\cdot)\}$, then for all $s < n$, the optimal decision is to stay idle since $\max\{V^0(s), V^1(s)\} = V^0(s)$, and for all $s > n$, the optimal decision is to attack the channel since $\max\{V^0(s), V^1(s)\} = V^1(s)$. This concludes our proof.

## APPENDIX B
### PROOF OF PROPOSITION 1

In order to demonstrate this proposition, we need to resolve the full balance equation under threshold policy $n$ at each state $j$:
$$u^n(j) = \sum_{i=0}^{+\infty} pt^n(i \to j)u^n(i) \tag{24}$$

where $pt^n(i \to j)$ denotes the transitioning probability from the state $i$ to state $j$ under threshold policy $n$. After some computations, we obtain the desired result.

## APPENDIX C
### PROOF OF THEOREM 2

The Proof follows the same procedure in Theorem 5 in [28]. It relies essentially on showing that $\lambda(n)$ is increasing. For the sake of space, we prove only that $\lambda(\cdot)$ grows with $n$. Therefore, we first seek a closed-form expression of the intersection point $\lambda(n)$, we obtain:
- if $n = 0$:
$$\lambda(0) = \frac{a(c-b)(ab+cb+1-ac-2b)}{b(1-c)(1-b)((1-b+a)(1-c) + a(c-b)/b)} \tag{25}$$
- if $n > 0$:
$$\begin{aligned}\lambda(n) = &\frac{(c-b)}{(1-b)^2(1-b+a)(1-c)} \\ &\times \Big[n(1-c)(1-b+a)(1-b) - ab^n(c-b) \\ &+ (c-b)b + (1-b)^2\Big]\end{aligned} \tag{26}$$

**Lemma 2.** *The sequence $\lambda(n)$ is strictly increasing with $n$.*

*Proof:* To prove this lemma, we use the fact that $c - b$ is positive. Indeed we have:
- For AoI-based model: $c - b = qp \geq 0$
- For AoII-based model: $c - b = pq(1-r) \geq 0$

Leveraging that and given that $(1-b)^2(1-b+a)(1-c)$ is positive, and $n(1-c)(1-b+a)(1-b)$ and $-ab^n(c-b)$ are increasing functions with $n$ since $(1-c)(1-b+a)(1-b)$ is positive and $b \leq 1$, then $\lambda(n)$ is increasing with $n$ when $n \geq 1$. By comparing $\lambda(0)$ and $\lambda(1)$, we find that $\lambda(0) \leq \lambda(1)$ using the fact that $r \leq 1/2$. ∎

The next steps follow the same line of the proof of Theorem 5 in [28].